\documentclass[twocolumn,prl,amsmath,amssymb,nofootinbib,superscriptaddress]{revtex4}

\usepackage{dcolumn}
\usepackage{amsmath}
\usepackage{graphicx}
\usepackage{rotating}
\usepackage{amsfonts}
\usepackage{amssymb}
\usepackage[hang]{subfigure}

\usepackage{graphicx}
\usepackage{psfig,color}
\usepackage{epsfig}

\begin{document}

\preprint{DESY 05-224}

\title {Photon Regeneration from Pseudoscalars at X-ray Laser Facilities} 

\author{Raul Rabadan}
\email{rabadan@ias.edu}
\affiliation{Institute for Advanced Study, Einstein Drive, Princeton, NJ 08540} 
\author{Andreas Ringwald}
\email{andreas.ringwald@desy.de}
\affiliation{Deutsches Elektronen-Synchrotron DESY, Notkestra\ss e 85, D-22607 Hamburg, Germany}
\author{Kris Sigurdson}
\altaffiliation{Hubble Fellow}
\email{krs@ias.edu}
\affiliation{Institute for Advanced Study, Einstein Drive, Princeton, NJ 08540}                                                                                                                                 

\begin{abstract}
Recently, the PVLAS collaboration has reported an anomalously large rotation of the polarization 
of light in the presence of a magnetic field. As a possible explanation they consider the 
existence of a light pseudoscalar particle coupled to two photons. In this note, we propose a method 
of independently testing this result by using a high-energy photon regeneration experiment 
(the X-ray analogue of ``invisible light shining through walls'') using the synchrotron X-rays 
from a free-electron laser (FEL).  With such an experiment the region of parameter space implied by 
PVLAS could be probed in a matter of minutes.
\end{abstract}


\maketitle

Many models beyond the Standard Model predict the existence of new very light pseudoscalar
particles which are very weakly coupled to ordinary matter. Such light particles would arise
if there was a global continuous symmetry in the theory that is spontaneously broken in the vacuum
--- a notable example being the axion~\cite{Weinberg:1978ma}
 arising from the breaking of a U(1) 
Peccei-Quinn symmetry~\cite{Peccei:1977hh}, 
introduced to explain the absence of strong $CP$ violation.

Such pseudoscalars couple to two photons via 
\begin{equation}
{\cal L}_{\phi \gamma \gamma} = - \frac{1}{4}\, g\, \phi\, F_{\mu \nu}\tilde{F}^{\mu \nu} = g\, \phi\, \vec{E}\cdot \vec{B} , 
\label{coupling}
\end{equation}
where $g$ is the pseudoscalar-photon coupling, $\phi$ is the field corresponding to the pseudoscalar, 
and $F_{\mu\nu}$ ($\tilde{F}^{\mu\nu}$) is the (dual) electromagnetic field strength tensor. Correspondingly, 
in the presence of a magnetic field $\vec B$, a photon of frequency $\omega$ may oscillate into a pseudoscalar 
particle of small mass $m_\phi < \omega$, and vice versa.     
The exploitation of this result is the basic idea behind photon regeneration~\cite{regeneration,VanBibber:1987rq} 
(sometimes called ``invisible light shining through walls'' experiments). Namely, if a beam of light with $N_0$ 
photons is shone across a magnetic field, a fraction of these photons will turn into pseudoscalars.  
This pseudoscalar beam can then propagate freely through a wall or another obstruction without being absorbed,  
and finally another magnetic field located on the other side of the wall can transform some of these pseudoscalars 
into $N_f$ photons --- apparently regenerating these photons out of nothing. 
This type of experiment was carried out in Brookhaven using two prototype magnets for the 
Collider Accelerator Beam and was used to exclude values of the pseudoscalar-photon coupling 
$g< 6.7  \times 10^{-7}\ {\rm GeV}^{-1}$ for $m_\phi < 10^{-3}$ {\rm eV}~\cite{Cameron:1993mr}.   

Recently the PVLAS collaboration has reported an anomalous signal in 
measurements of the rotation of the polarization of photons in a magnetic field~\cite{Zavattini:2005tm}. 
One \emph{a priori} possible explanation of this apparent vacuum magnetic dichroism is the 
production of a pseudoscalar coupled to photons through Eq.~(\ref{coupling}), according to which
photons polarized parallel to the magnetic field disappear, leading to a rotation of the 
polarization plane~\cite{Maiani:1986md}. The region quoted in Ref.~\cite{Zavattini:2005tm} that might explain 
this signal is  $1.7 \times 10^{-6}\  {\rm GeV}^{-1}<  g < 1.0 \times 10^{-5}\  {\rm GeV}^{-1}$ for 
$0.7 \times  10^{-3}\  {\rm eV} < m_\phi < 2.0 \times  10^{-3}\  {\rm eV}$, obtained from a combination of
previous limits on $g$ vs. $m_\phi$ from a similar, but less sensitive polarization experiment in 
Brookhaven~\cite{Cameron:1993mr} and the $g$ vs. $m_\phi$ curve corresponding to the PVLAS signal.  
A pseudoscalar-photon coupling in this region of parameter space is in contradiction with limits derived from  
pseudoscalar production in stars~\cite{Raffelt:1985nk}, particularly in the sun~\cite{Andriamonje:2004hi,Raffelt:2005mt}.  
However, in principle one can try to find some 
non-minimal models where pseudoscalar production in stars is small~\cite{Masso:2005ym} to resolve the discrepancy 
with the laboratory result.  

The main motivation of this note is to suggest an independent laboratory probe 
of the $g \phi \vec{E}\cdot\vec{B}$ interaction without reference to axion production in stars 
(see~\cite{Dupays,Kleban:2005rj}).  Given the unexpected and surprising results found in the neutrino sector 
we believe this type of laboratory cross-check is certainly warranted. 
Specifically, we consider the possibility of exploiting a powerful X-ray free-electron laser in a photon 
regeneration experiment\footnote{This idea has been considered first in Ref.~\cite{Ringwald:2001cp}, 
with similar sensitivity estimates as in the present note.}  
to probe the region where the PVLAS signal could be explained in terms of a light pseudoscalar 
particle.\footnote{For the proposal of another photon regeneration experiment exploiting 
an ordinary optical laser to test PVLAS, see Ref.~\cite{LoI}.} 

Two facilities have designed and, in fact, are about to commence construction of powerful free-electron lasers (FEL) in 
the X-ray range: the Linac Coherent Light Source (LCLS) at SLAC~\cite{lcls} and the European X-Ray Laser  XFEL at DESY~\cite{xfel}. 
The LCLS is a free-electron X-ray laser that will use the last kilometer of the SLAC linear accelerator.  
It will be capable of producing intense pulses of X-ray photons at energies between 0.8 keV and 8 keV. 
Project completion is expected in 2008 and the first experiments involving the LCLS will be running in 2009.  
The XFEL at DESY starting in 2012 will have several lasers with similar characteristics with photon energies 
in the 1-10 keV range and an average flux of photons of approximately  $10^{17}$--$10^{19}$ photons per second.
Already running at the DESY TESLA Test Facility is an FEL which provides tunable radiation from the vacuum-ultraviolet (10 eV)
to soft X-rays (200 eV), with an average flux of about $10^{18}$--$10^{19}$ photons per second~\cite{vuvfel}.  

%%%%%%%%%%%%%%%%%%%%%%%%
\begin{figure}[t]
\begin{center}
\epsfxsize=3.5in
\epsffile{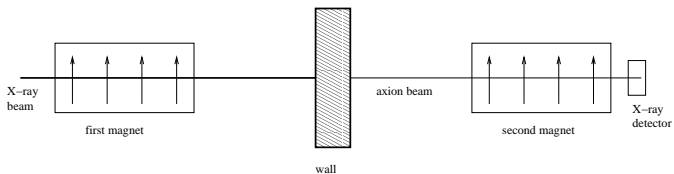}
\end{center}
\caption{\small Schematic figure of the regeneration experiment.} 
\label{regeneration} 
\end{figure}
%%%%%%%%%%%%%%%%%%%%%%%%%

Our benchmark proposal uses a photon regeneration set up with two equal magnets of magnetic field $B$ and length $L$.  
The first of them converts the X-ray photons from the laser beam into pseudoscalars and the second, on the other side 
of the ``wall'',  converts the high-energy pseudoscalars into X-ray photons again (see Fig.~\ref{regeneration}). 
We consider the sensitivity of two experimental setups: a superconducting magnet of $L=10$~m and $B= 10$ T and 
a conventional magnet with $L=20$~m and $B= 1$ T.  The first setup is more appropriate for the DESY FEL facilities  
because of the availability of superconducting magnets after the decommissioning of the 
electron-proton collider HERA in mid of 2007~\cite{Ringwald:2003ns}.

The probability of photon-pseudoscalar conversion in a constant magnetic field of length $L$ is:
\begin{equation}
P = \frac{1}{4} g^2B^2L^2j_0^2\left(\frac{qL}{2}\right) = g^2 B^2 \frac{\sin^2{\left(\frac{q L}{2}\right)}}{q^2} \, , 
\end{equation}
where $q = \omega - \sqrt{\omega^2 -m_{\phi}^2}$ is the difference between the momentum of the pseudoscalar and 
the photon.  When the mass of the pseudoscalar is much smaller than the photon energy, we can approximate 
$q = m_{\phi}^2/2\omega$. For the magnets and pseudoscalar masses we are considering, we have  $qL\ll 1$,  so that 
$j_{0} \rightarrow 1$ and the conversion probability simplifies to\footnote{\label{coherence}In Ref.~\cite{VanBibber:1987rq}
it was claimed incorrectly~\cite{vanBibber:private} that the use of a light source of coherence length $\ell_c$
degrades the conversion probability by a factor of $\ell_c/L$, if $\ell_c<L$. 
This lead to a substantial underestimate of the XFEL sensitivity for photon regeneration  
in Ref.~\cite{Ringwald:2003ns}.}
\begin{equation}
P = \frac{1}{4}g^2 B^2 L^2  \, . 
\end{equation}
Using magnets of a length of $10~ {\rm m} = 5.07 \times 10^7\ {\rm eV}^{-1}$ and a magnetic field of 
$10~{\rm T} = 1.95 \times 10^{3}\ {\rm eV}^2$, the probability of converting a photon into a pseudoscalar is
\begin{equation}
P = 2.4 \times 10^{-9} \left( \frac{g}{10^{-6}\ {\rm GeV}^{-1}} \right)^2 
\left( \frac{B}{10\ {\rm T}}\right)^2  \left( \frac{L}{10\ {\rm m}}\right)^2 . 
\end{equation}

%%%%%%%%%%%%%%%%%%%%%%%%
\begin{figure}
\begin{center}
\epsfxsize=3.33in
\epsffile{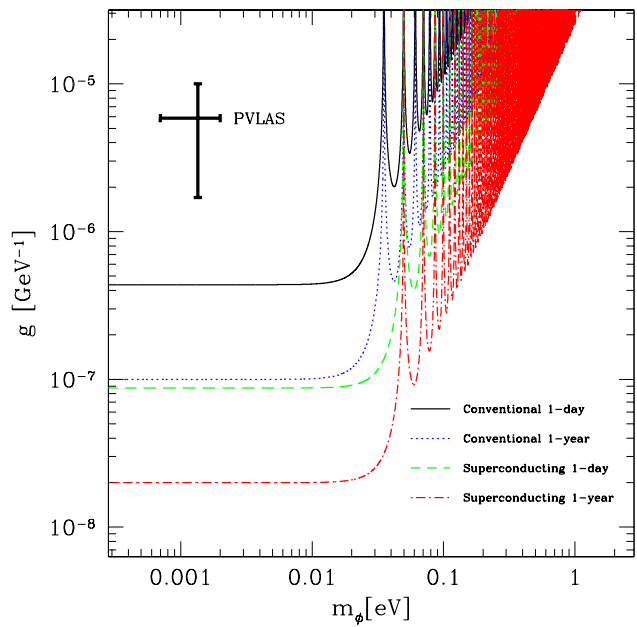}
\end{center}
\caption{\small 95\% confidence level exclusion region for different running times: 1 day and 1 year for two 
different magnets (a conventional magnet with $L=20$~m and $B=1$~T and a superconducting magnet with $L=10$~m and $B=10$~T).  
We have assumed efficient X-ray detection and an X-ray beam with $N_{0} = 10^{17}\ {\rm s}^{-1}$ and $\omega = 10$~keV.} 
\label{bounds} 
\end{figure}
%%%%%%%%%%%%%%%%%%%%%%%%%

An X-ray laser facility may produce on average as many as $N_{0} \simeq {\cal N}_{17} \times 10^{17}$ photons per second 
(where ${\cal N}_{17} \simeq 1$-$100$). The number of pseudoscalar particles that are produced per second is 
$N_\phi = P N_{0}$. 
The number that will be transformed back into photons is just $N_f = P N_{\phi} = P^2 N_{0}$.  
Thus we find the photon regeneration rate as
\begin{equation}
N_f = 0.6 {\rm s}^{-1} {\cal N}_{17}\left( \frac{g}{10^{-6}\ {\rm GeV}^{-1}} \right)^4 
\left( \frac{B}{10\ {\rm T}}\right)^4  \left( \frac{L}{10\ {\rm m}}\right)^4 \, . 
\end{equation}

We immediately see that the PVLAS result can be tested in a
matter of minutes. We summarize the full mass dependence of these potential bounds and the values 
of the pseudoscalar-photon coupling that can be probed in this way for various running times in Fig.~\ref{bounds}.  
For the superconducting magnet and single day experiment, the region $g > 8.9 \times 10^{-8}\ {\rm GeV}^{-1}$ 
could be probed at 95\% confidence, while in a year the limit could be improved to $g > 2.0 \times 10^{-8}\ {\rm GeV}^{-1}$. 

Let us summarize why we believe our proposal is of interest. Firstly, high frequency photons are able to avoid 
pseudoscalar-photon incoherent effects, which set in for $m_\phi^2 L\approx 2\pi\omega$, 
and can probe pseudoscalar masses\footnote{Higher masses can be probed 
by filling in buffer gas into the magnetic field region~\cite{Andriamonje:2004hi}.} $m_\phi < 0.05$ eV, beyond the 
capability of optical photons.  Secondly, X-ray detection 
can be very efficient, with an efficiency $\eta \simeq 1$, and backgrounds in the laboratory are very much under 
control.  They can be reduced by utilizing the directionality of the signal and comparing the backgrounds when the 
magnets are on with the background when the magnets are off.  Thirdly, superconducting 
magnets are not mandatory and so, in principle, one can use long ordinary magnets that are more cost effective.  
Moreover,  a remarkable feature of this proposal is that, if the 
initial conversion magnet is placed before a target that is the subject of other experiments, it is possible to 
perform both experiments simultaneously, since the pseudoscalar beam will propagate unimpeded through the target.

This experiment could probe the region of parameter space relevant to PVLAS in less than a day.
Moreover, it could serve as a test facility for an ambitious large scale photon regeneration 
experiment~\cite{Ringwald:2003ns}$^{\ref{coherence}}$,  based on the XFEL and the recycling of HERA's  $400$ 
superconducting dipole magnets ($B=5$~T, $L=200\times 10$~m), to reach within a year an unprecedented sensitivity 
of $g > 2.0 \times 10^{-10}\ {\rm GeV}^{-1}$, comparable to the 
limits involving model-dependent astrophysical considerations~\cite{Raffelt:1985nk,Andriamonje:2004hi,Raffelt:2005mt}.  

\vspace{0.15cm}
\centerline{\bf Acknowledgments}
\vspace{0.05cm}
We would like to thank Karl van Bibber, Matthew Kleban, 
Eduard Masso, Carlos Pe\~na Garay, Pierre Sikivie, Thomas Tschentscher, 
and Daniel Q. Wang for useful discussions.  
RR is supported by DOE grant DE-FG02-90ER40542. KS is supported by NASA through Hubble 
Fellowship grant HST-HF-01191.01-A awarded by the Space Telescope
Science Institute, which is operated by the Association of Universities for Research in Astronomy, 
Inc., for NASA, under contract NAS 5-26555.

\end{document}